\documentclass{iau}
\usepackage{graphicx}

\title[Nonthermal Filaments] 
{Nonthermal Filamentary Radio Features Within 20 pc of the Galactic Center}

\author[M.\ R.\ Morris, J.-H.\ Zhao \& W.\ M.\ Goss]   
{Mark R.\ Morris$^1$, Jun-Hui Zhao$^2$ \and W. M. Goss$^3$}

\affiliation{$^1$Dept. of Physics \& Astronomy, University of California, Los Angeles, CA 90095-1547, USA\\ email: {\tt morris@astro.ucla.edu} \\[\affilskip]
$^2$Harvard-Smithsonian CfA, 60 Garden Street, MS 78, Cambridge, MA 02138, USA \\[\affilskip]
$^3$NRAO, P.O. Box O, Socorro, NM 87801, USA}

\pubyear{2014}
\volume{303}  
\jname{The Galactic Center: Feeding and Feedback in a Normal Galactic Nucleus}
\editors{L. Sjouwerman, C.C. Lang \& J. Ott, eds.}
\begin{document}

\maketitle

\begin{abstract}
Deep imaging of the Sgr A complex at 6 cm wavelength with the B and C configurations of the Karl G. Jansky VLA\footnote {The Jansky Very Large Array (JVLA) is operated by the National Radio Astronomy Observatory
(NRAO). The NRAO is a facility of the National Science Foundation operated under
cooperative agreement by Associated Universities, Inc.} has revealed a new population of faint radio filaments.  Like their brighter counterparts that have been observed throughout the Galactic Center on larger scales, these filaments can extend up to $\sim$10 parsecs, and in most cases are strikingly uniform in brightness and curvature. Comparison with a survey of Paschen-$\alpha$ emission reveals that some of the filaments are emitting thermally, but most of these structures are nonthermal: local magnetic flux tubes illuminated by synchrotron emission.  The new image reveals considerable filamentary substructure in previously known nonthermal filaments (NTFs). 
Unlike NTFs previously observed on larger scales, which tend to show a predominant orientation roughly perpendicular to the Galactic plane, the NTFs in the vicinity of the Sgr A complex are relatively randomly oriented. 
Two well-known radio sources to the south of Sgr A -- sources E and F -- consist of numerous quasi-parallel filaments that now appear to be particularly bright portions of a much larger, strongly curved, continuous, nonthermal radio structure that we refer to as the ÓSouthern CurlÓ. It is therefore unlikely that sources E and F are HII regions or pulsar wind nebulae. The Southern Curl has a smaller counterpart on the opposite side of the Galactic Center -- the Northern Curl -- that, except for its smaller scale and smaller distance from the center, is roughly point-reflection symmetric with respect to the Southern Curl.  The curl features indicate that some field lines are strongly distorted, presumably by mass flows. The point symmetry about the center then suggests that the flows originate near the center and are somewhat collimated. 
\end{abstract}

\firstsection 
\section{Introduction}

Since their discovery about 30 years ago by \cite{YZMC84}, the population of nonthermal radio filaments (NTFs) has emerged as one of the unique characteristics of the Galactic center environment.  Numerous NTFs have been examined across the radio spectrum, and it has become clear that they are magnetically organized structures in which the magnetic field is parallel to the filaments, and that their illumination in the radio is owed to locally injected relativistic electrons (Reich 1994; Lang et al.\ 1999a).  They are typically quite linear, showing only gentle curvature on scales up to tens of parsecs, which led Yusef-Zadeh \& Morris (1987a, see also Morris \& Serabyn 1996) to suggest that the field within them must be quite rigid, or strong ($\sim$milligauss) in order to withstand the distorting effects of the velocity dispersion of the interstellar medium, especially where they interact with dense, turbulent molecular clouds. From an early stage, it was noted that many of the brightest NTFs are oriented perpendicular to the Galactic plane (Morris 1994, 1996; LaRosa et al.\ 2004; Nord et al.\ 2004), which led to the suggestion that a uniform vertical field permeates the Galactic center (GC).  However, several studies have also shown that many of the weaker, shorter NTFs are essentially oriented at random (Lang et al. 1999b; Yusef-Zadeh 2003; LaRosa et al.\ 2004), which has raised the question of whether some other, local phenomenon is responsible for the filament orientations (Shore \& LaRosa 1999; Bicknell \& Li 2001; Yusef-Zadeh 2003; Boldyrev \& Yusef-Zadeh 2006).   A compilation of over 80 NTFs observed throughout the GC at 20cm with the VLA emphasized the trend that most of the large, bright NTFs are roughly perpendicular to the Galactic plane, but showed as well that many of the small NTFs do not conform to that trend (Yusef-Zadeh et al.\ 1984, see their fig.\ 29).

\begin{figure}[t]
\begin{center}
 \includegraphics[width=3in]{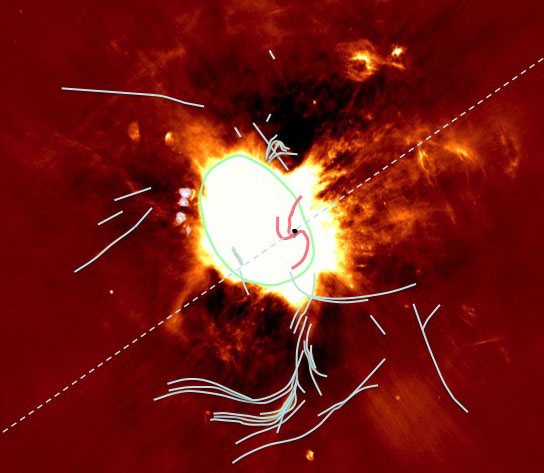} 
 \caption{Diagram showing the traces of all the detected NTFs surrounding Sgr A, superimposed on our 6cm JVLA radio image.  The outline of Sgr A East and the thermal arms of Sgr A West are also sketched upon the saturated portion of the radiograph. The black
 dot indicates the location of Sgr A*.  North is up and East is to the left.  The vertical to the Galactic plane is indicated as a dashed line. In all figures in this paper, the beam size is 1.6'' $\times$ 0.6'', with major axis position angle of 12$^{\circ}$. The angular width of the region shown is 13.6 arcmin, corresponding to 40 pc at the Galactic center distance of 8 kpc.}
   \label{fig1}
\end{center}
\end{figure}

\section{Radio Filaments Near Sgr A}

In previous radio studies of the immediate region around Sgr A, several filamentary structures have been noted (Yusef-Zadeh \& Morris 1987b; Yusef-Zadeh et al.\ 2004).  In the deep imaging that we have recently done with the JVLA (see Zhao et al.\ 2014, and in this volume, for a description of the observations), we have identified many new filamentary features, most of which are NTFs, but a number of them are apparently thermal features, since they show Paschen-$\alpha$ counterparts in the survey images produced by Wang et al. (2010).  Polarization and spectral index measurements -- both in progress on this same data set $-$ will provide additional constraints on the nature of the observed filaments.  A schematic diagram of the locations of all the detected NTF's is shown in Fig.\,\ref{fig1}.  As is frequently the case with the NTFs, many of these consist of multiple closely-spaced parallel subfilaments.  

Although several of the filaments are roughly perpendicular to the Galactic plane (parallel to the dashed line in Fig.\ 1), there is no obvious trend to their orientations.  This is particularly true for the shorter NTFs, consistent with previous studies on larger scales.  The surface density of filaments is rather high compared with that reported in previous studies of any location in the GC, probably because of the very high sensitivity ($\sim$10 $\mu$Jy rms) and dynamic range (80,000) achieved with our JVLA imaging.  

\begin{figure}[b]
\begin{center}
 \includegraphics[width=4in]{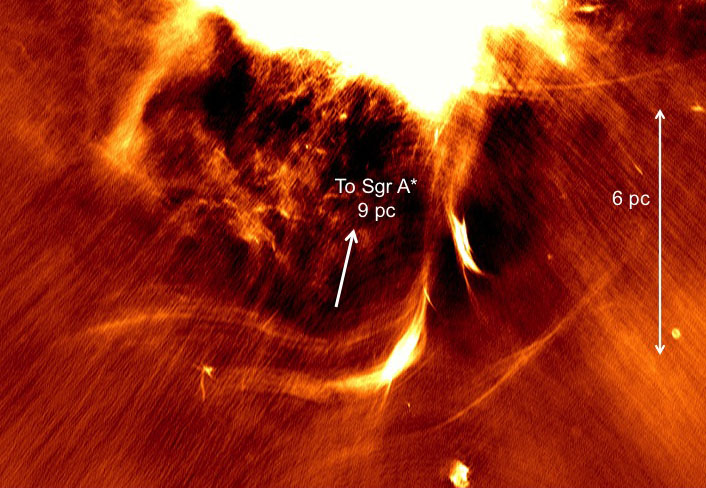} 
 \caption{The``Southern Curl'': a continuous magnetic structure, possibly resulting from the impact
 of a collimated outflow from near Sgr A*.  Several other NTFs are evident in this image.  Background
 striations oriented radially with respect to Sgr A* are artifacts of incompletely corrected side lobes. The
 angular width of the region shown is 6.8', corresponding to 17 pc.}
   \label{fig2}
\end{center}
\end{figure}

\begin{figure}[t]
\begin{center}
 \includegraphics[width=3.2in]{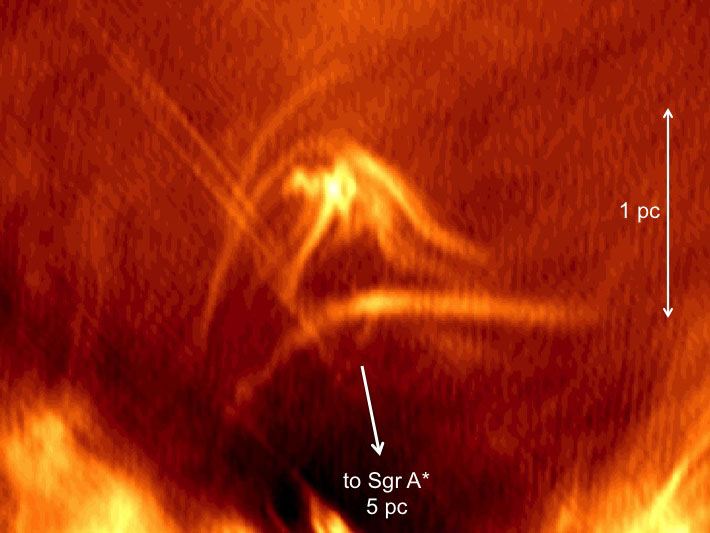} 
 \caption{The ``Northern Curl'': bent magnetic field lines, possibly resulting from the impact
 of a collimated outflow from near Sgr A*.  This feature is crisscrossed by a split pair of linear NTFs
 with which it appears to be interacting; note the discontinuity in the most prominent filament of 
 the Northern Curl at the position where it meets one of the linear NTFs.  The angular width of the 
 region shown is 1.35 arcmin, corresponding to 3.4 pc}
   \label{fig3}
\end{center}
\end{figure}

\section{The "Curls" -- strongly deformed NTFs}

Two filamentary complexes observed in our study show unusually strong curvature -- the Northern and Southern ``Curls'' straddling the Sgr A complex.  
The brightest portions of the Southern Curl were identified as two radio continuum sources, Sgr A-E and F, by Ho et al. (1985).   \cite{YZM87b} noted that they are parts of large-scale "semi-circular features".  The radio and X-ray emission properties of sources E (G359.88-0.08) and F (G359.90-0.06) were examined in detail by Yusef-Zadeh et al.\ (2005).   Our new 6cm image (Fig.\,\ref{fig2}) indeed shows not only that they are portions of extended, strongly curved structures, but also that those structures, and sources E and F within them, consist of bundles of continuous nonthermal filaments.  We conclude that the entire structure of the Southern Curl is a continuous, organized magnetic field structure illuminated along its length by synchrotron emission, and that sources E and F are locations where the energies and/or densities of the relativistic electrons are particularly high.  The morphology of these sources is inconsistent with their being either portions of supernova remnants or pulsar wind nebulae.

The Northern Curl, shown in Fig.\,\ref{fig3}, has a similar morphology, although it is a smaller structure.  It had previously been identified as sources I1 and I2 by \cite{YZM87b} and as object A7 by \cite{YZHC04}, but its intricate filamentary structure has not previously been revealed.  We note that it lies on the opposite side of Sgr A* from the Southern Curl, and except for its smaller scale and closer proximity to Sgr A*, is almost a point symmetric counterpart to the Southern Curl.  

A line joining the positions of maximum curvature of the Northern and Southern Curls passes close to Sgr A*.  This spatial relationship leads us to hypothesize that a collimated outflow of plasma from Sgr A* or its near vicinity is responsible for distorting the ambient magnetic field lines
(which might otherwise have been vertical) at these two locations.  The shocked impact of such an outflow could also be responsible for generating the relativistic electrons via diffusive shock acceleration, which could account for the brightest portions of the curls being located close to the positions of maximum curvature.

\end{document}